\documentclass[%
 sor,
 jor,
 amsmath,amssymb,
 reprint,%
longbibliography
]{revtex4-1}

\usepackage{graphicx}% Include figure files
\usepackage{dcolumn}% Align table columns on decimal point
\usepackage{bm}% bold math

\usepackage{color}
\definecolor {myc} {rgb} {0,0,0}  %black, uncomment this line for regular copy

\newcommand{\um} {~\mathrm{\mu m}}

\begin{document}

\preprint{AIP/123-QED}

\title[Power law viscoelasticity of a fractal colloidal gel]{Power law viscoelasticity of a fractal colloidal gel}

\author{S. Aime}
 \email{stefano.aime@gmail.com.}
\author{L. Cipelletti}
 \email{Luca.Cipelletti@umontpellier.fr.}
 \author{L. Ramos}%
 \email{Laurence.Ramos@umontpellier.fr.}
\affiliation{Laboratoire Charles Coulomb (L2C), Univ. Montpellier, CNRS, Montpellier, France}%

\date{\today}

\begin{abstract}

Power law rheology is of widespread occurrence in complex materials that are characterized by the presence of a very broad range of microstructural length and time scales. Although phenomenological models able to reproduce the observed rheological features exist, in general a well-established connection with the microscopic origin of this mechanical behavior is still missing. As a model system, this work focuses on a fractal colloidal gel. We thoroughly characterize the linear power law rheology of the sample and its age dependence. We show that at all sample ages and for a variety of rheological tests the gel linear viscoelasticity is very accurately described by a Fractional Maxwell (FM) model, characterized by a power law behavior. Thanks to a unique set-up that couples small-angle static and dynamic light scattering to rheological measurements, we  {\color {myc} show that in the linear regime shear induces reversible non-affine rearrangements which might be at the origin of the power law rheology} and we discuss the possible relationship between the FM model and the microscopic structure of the gel.

\end{abstract}

\pacs{XXX}
\keywords{XXX}

\maketitle

\section{Introduction}

	Power law rheology is of widespread occurrence in complex materials that do not exhibit one unique relaxation time, from biomaterials, such as cells \cite{fabry_scaling_2001, djordjevic_fractional_2003, desprat_creep_2005, balland_power_2006, kollmannsberger_linear_2011, hecht_imaging_2015}, tissues \cite{kohandel_frequency_2005, davis_constitutive_2006, shen_fractional_2013}, or biopolymer networks \cite{gobeaux_power_2010, curtis_validation_2015} and pastes \cite{jozwiak_fractional_2015}, to food science \cite{ma_simulating_1996, zhou_effect_1998, subramanian_linear_2006, ng_power_2008, caggioni_rheology_2007, korus_impact_2009, moreira_rheology_2011, ronda_impact_2013, xu_fractional-order_2013, jaishankar_fractional_2014, leocmach_creep_2014-1, jozwiak_fractional_2015, faber_describing_2017, faber_describing_2017-1}, colloidal gels \cite{rich_size_2011}, microgels \cite{lidon_power-law_2017}, hydrogels \cite{hung_fractal_2015}, polymer gels \cite{chambon_rheology_1986, winter_analysis_1986, durand_frequency_1987, martin_viscoelasticity_1988, adolf_time-cure_1990, tirtaatmadja_superposition_1997, larsen_microrheology_2008, leibler_dynamics_1991-1}, melts \cite{plazek_dynamic_1960, hernandez-jimenez_relaxation_2002, friedrich_constitutive_1999}, elastomers \cite{curro_theoretical_1983, h.h._winter_rheology, john_d._ferry_viscoelastic_1980}, and composites \cite{metzler_relaxation_1995}.
	Although the generality of power law rheology is captured by phenomenological models such as the Soft Glassy Rheology model \cite{sollich_rheology_1997, sollich_rheological_1998}, in general a well established connection with the microscopic origin of this mechanical behavior is still missing. Few exceptions include the parallel drawn between power law rheology and microscopic structure and dynamics in the framework of polymer physics. This is for example the case of Rouse motion, where the self-similar relaxation dynamics naturally come into play as a consequence of the fractal nature of the polymer coil \cite{ralph_h._colby_polymer_2003}. In this case, the fractal dimension $d_f=2$ describing the microscopic structure is directly linked to the fractional order of the rheological model, i.e. the exponent $\alpha=0.5$ describing the power law rheology.
	Similarly, in the attempt of modeling polymer gels at the critical point, Adolf and Martin \cite{adolf_time-cure_1990} explicitly derive a power law distribution of relaxation times from a postulated scale-independence, which is assumed to hold at the critical gel point according to percolation theory \cite{stauffer_introduction_2003}, in quantitative agreement with a direct measurement of the fractal mass distribution \cite{james_e._martin_sol-gel_1991}.
	More generally, a quantitative link between power law rheology and the microscopic structures of critical gels with arbitrary fractal dimension is established by Muthukumar \cite{muthukumar_screening_1989} in the two limit cases of screened and unscreened excluded volume and hydrodynamic interactions. This model essentially extends Rouse dynamics from a single chain to a branched, fractal object. In this case the mathematical formulation is much more involved, but, as in Rouse dynamics, the power law rheology reveals the self-similarity of microscopic dynamics, stemming from the fractal structure {\color{myc} \cite{patricio_rheology_2015}}. Accordingly, the direct link between fractal structure and power law rheology has proved to properly describe several experimental results on various polymer networks \cite{dahesh_spontaneous_2016, takenaka_comparison_2004, matsumoto_viscoelastic_1992}.
	Although fractal structures are very well represented in soft materials, from human tissues \cite{helmberger_quantification_2014, mauroy_optimal_2004, ahmadi_brain_2013} to polymer and colloidal gels \cite{james_e._martin_sol-gel_1991, carpineti_transition_1993, bremer_fractal_1990}, the wide spectrum of systems displaying a power law rheology suggests that power law rheology does not necessarily stem from fractal structure, and that other microscopic mechanisms might produce power law distributions of relaxation times as well. In the absence of a full understanding of such mechanisms, however, fractional rheological models {\color {myc} that consider fractional derivatives of the stress and/or the strain in the constitutive equations} remain largely phenomenological, and the amount of relevant physical information that can be extracted from them remains arguable.
	
	In this regard, one crucial aspect is the scarcity of systems for which a thorough discussion of linear rheology is available. Indeed, the majority of the works finding a power law rheology only focuses on oscillatory shear \cite{ma_simulating_1996, metzler_relaxation_1995, tirtaatmadja_superposition_1997, fabry_scaling_2001, djordjevic_fractional_2003, shen_fractional_2013, caggioni_rheology_2007, rich_size_2011, hung_fractal_2015, friedrich_constitutive_1999}, whereas others only focus on transient experiments, either stress relaxation \cite{hernandez-jimenez_relaxation_2002, curro_theoretical_1983}, creep \cite{desprat_creep_2005, xu_fractional-order_2013} or both \cite{davis_constitutive_2006}.
	Transient experiments alone are delicate, and have to be carefully performed in order to ensure that only the linear regime is being probed, since sometimes power law creep clearly emerges as a nonlinear phenomenon \cite{duval_creep_2010, paredes_rheology_2013}. Indeed, plastic power law creep is observed for many systems, from crystalline materials like metals \cite{e.n._andrade_viscous_1910, miguel_dislocation_2002, jean-paul_poirier_creep_1985, cottrell_time_1952} or ice \cite{ashby_creep_1985} to colloidal \cite{siebenburger_creep_2012, sentjabrskaja_creep_2015, coussot_aging_2006, caton_plastic_2008} or polymeric \cite{nechad_creep_2005, karobi_creep_2016} systems, which would have a very different rheology in the linear regime.
	On the other hand, the challenge of oscillatory shear is that it typically gives access to a limited range of frequencies, which sometimes makes it difficult to distinguish the predictions of different models. Moreover, measurements are even more difficult for non-stationary, aging, systems, although clever techniques exist to overcome those limitations \cite{ghiringhelli_optimal_2012, bouzid_computing_2018, geri_time-resolved_2018}. In some systems a larger spectrum can be accessed exploiting time temperature superposition, although its applicability is far from being trivial in most systems cited above.

	One instructive example in this regard is represented by various independent works on flour doughs \cite{korus_impact_2009, moreira_rheology_2011, ronda_impact_2013}, where oscillatory rheology revealed clear power law responses, whereas creep and recovery on the same samples were nicely fitted by a Burger model, which predicts an exponential creep. In another example on polymer microgels, a power law rheology was found in both oscillatory shear and creep, but described by incompatible power law exponents \cite{lidon_power-law_2017}, which also questions the applicability of a fractional model. Moreover, for polymer melts, a power law creep was also interpreted as the short time limit of a stretched exponential \cite{cheriere_three_1997}.
	All these examples show that a comparison between different rheological measurements is essential to truly validate a phenomenological model. This is done for example on individual cells \cite{balland_power_2006}, but also on collagen networks \cite{gobeaux_power_2010}, biopolymer gels \cite{ng_power_2008, leocmach_creep_2014-1}, natural gums \cite{jaishankar_fractional_2014} and cheese \cite{faber_describing_2017-1}. For these systems, fractional rheological models provide a consistent description of the observed rheology.
	However, most of these systems are rather complex, and their structural analysis is delicate, rendering virtually difficult clear connections between rheological and structural properties.
	
\bigbreak

	In this paper we focus on a simpler model system, namely a colloidal gel, for which both microscopic structure and spontaneous dynamics are well known and can be easily measured with scattering methods \cite{carpineti_mass_1995, cipelletti_universal_2000}. Thanks to a careful analysis of both oscillatory and transient shear we show that the linear rheology is consistently described by a Fractional Maxwell model, whose parameters are discussed in detail, with reference to the sample structure and to its aging properties. By coupling rheology to light scattering, we moreover demonstrate that, in the linear viscoelastic regime, the shear deformation provokes microscopic rearrangements that are fully reversible. The paper is organized as follows: after reviewing the theoretical background (Sec.~\ref{sec:theory}), we describe the sample preparation and the experimental setup (Sec.~\ref{sec:matmeth}). In Sec.~\ref{sec:experimental} the rheological results are shown and fully described with the linear viscoelastic model, and the link to structural data is established thanks to a set-up that couples small-angle light scattering to rheological measurements. A conclusive section (Sec.~\ref{sec:discussion}) closes the manuscript, with a detailed discussion about the link between the rheological properties and the microscopic structure.

\section{Theoretical background}
\label{sec:theory}

We recall here the theoretical basis for fractional rheology, which can be found in greater detail in Refs. \cite{rudolph_hilfer_applications_2000, jaishankar_power-law_2013}

\subsection{Fractional derivatives in rheology}

	For systems characterized by a power law rheology, the relaxation modulus measured, e.g., in a step strain experiment, decays as $G(t)\propto t^{-\alpha}$ with $0<\alpha<1$. For such systems, fractional expressions come naturally into play as a result of the superposition principle. Indeed, the stress response to an arbitrary shear history $\gamma(t)$ is given by the convolution integral \cite{rudolph_hilfer_applications_2000}:

%+++++++++++++++++++++++++
\begin{equation}
	\sigma(t) = \frac{K}{\Gamma(1-\alpha)} \tau^\alpha \int\limits_{-\infty}^t  (t-t')^{-\alpha}\dot\gamma(t') dt'
\end{equation}
%+++++++++++++++++++++++++

\noindent
Here $K$ is a modulus and $\tau$ is a characteristic time. In this expression one recognizes the fractional derivative of order $\alpha$, $\mathrm{d}^\alpha/\mathrm{d}t^\alpha$, which is defined as $\Gamma(1-\alpha) d^\alpha \gamma/dt^\alpha = \int\limits_{-\infty}^t  (t-t')^{-\alpha}\dot\gamma(t') dt'$ \cite{demirci_method_2012}. Hence,

%+++++++++++++++++++++++++
\begin{equation}
	\sigma(t)=K \tau^\alpha\frac{d^\alpha \gamma}{dt^\alpha}
	\label{eqn:springpot_constitutive}
\end{equation}
%+++++++++++++++++++++++++

\noindent
By inverting the above equation, one finds for example that the creep deformation of the material in response to a step stress applied at time $t=0$, $\sigma(t)=\sigma_0\Theta(t)$, with $\Theta(t)$ the Heaviside step function, is a power law:

%+++++++++++++++++++++++++
\begin{equation}
	\gamma(t)=K^{-1} \tau^{-\alpha} \frac{d^{-\alpha}}{dt^{-\alpha}}\sigma(t)=\frac{\sigma_0}{K}\frac{\Theta(t)}{\Gamma(1+\alpha)} \left(\frac{t}{\tau}\right)^\alpha
	\label{eqn:springpot_creep}
\end{equation}
%+++++++++++++++++++++++++

\begin{figure}[h]
\centering
  \includegraphics[width=1\columnwidth]{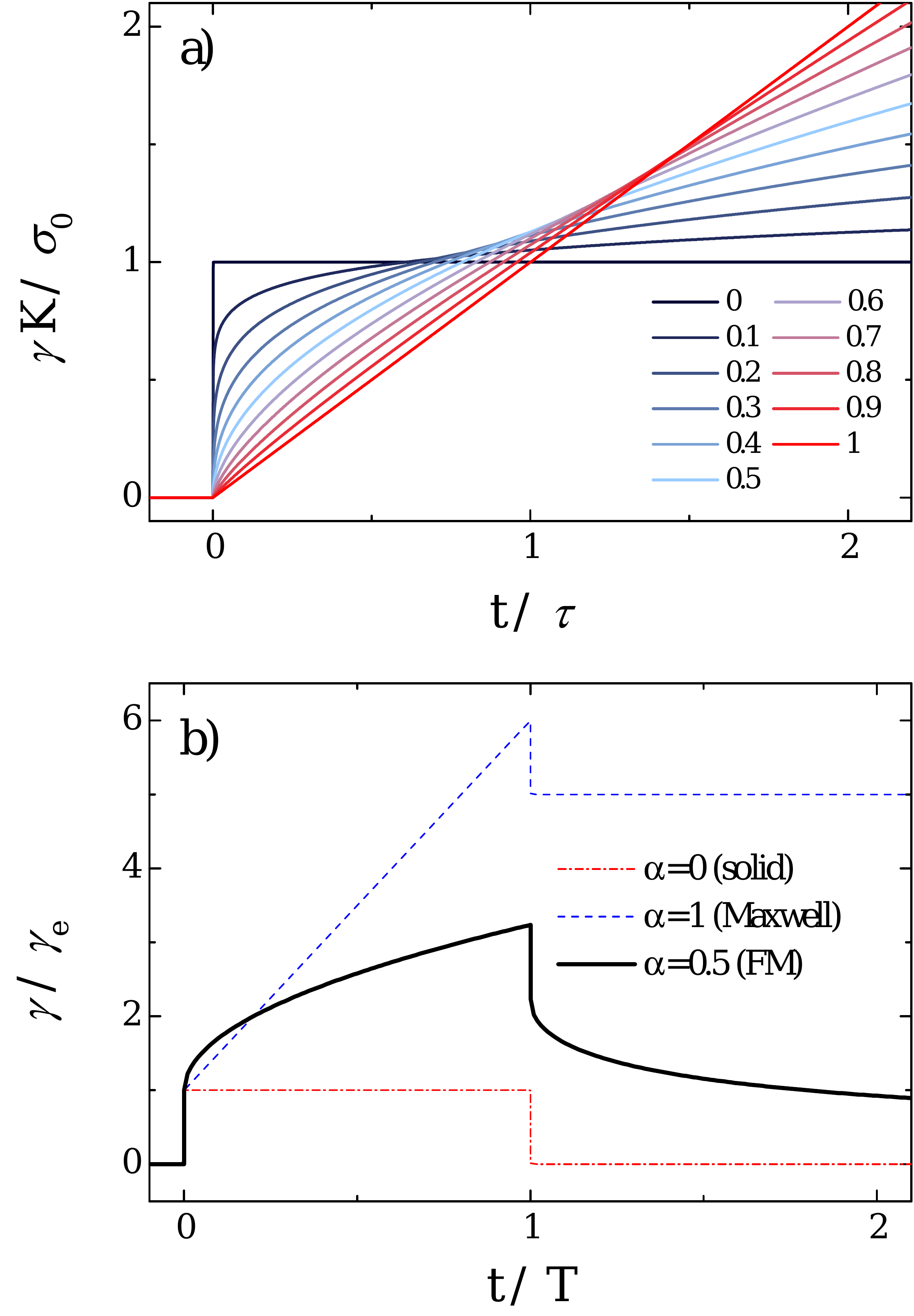}
\caption{a) Normalized strain as a function of the normalized time upon application of a constant stress of amplitude $\sigma_0$ applied at time $0$ for a springpot with different exponents $\alpha$ as indicated in the legend. b) Time evolution of the strain normalized by the initial elastic jump during the creep and recovery of an elastic solid ($\alpha=0$), a Maxwell fluid ($\alpha=1$) and a fractional Maxwell fluid with an exponent $\alpha=0.5$.}
  \label{fig:springpot}
\end{figure}

\noindent
For $\alpha=1$ one recovers the linear flow of a dashpot element, $\gamma(t)=\sigma_0 t / K \tau$ (for $t\geq0$), where $K\tau$ is a viscosity. On the other hand, for $\alpha=0$, one recovers the step strain of an elastic spring of elastic modulus $K$, $\gamma(t)=\sigma_0/K$. In the general case ($0<\alpha<1$), Eq.~\ref{eqn:springpot_creep} corresponds to a fractional rheological element called Scott-Blair element or springpot \cite{blair_limitations_1947, bagley_theoretical_1983, bagley_fractional_1986}. Figure~\ref{fig:springpot} (top) reproduces the creep profiles expected for springpot elements with different values of the fractional exponent $\alpha$ and shows a smooth and continuous transition form a viscous liquid to an elastic solid.

%------------------------------------------------
%------------------------------------------------

\subsection{Fractional Maxwell Model}
The Fractional Maxwell (FM) model is  based on two springpots in series, and is thus characterized by four independent parameters, which can be interpreted as an elastic modulus $G_0$, a characteristic time $\tau_{FM}$, and two exponents $\alpha$ and $\beta$.  In the following, we will consider the special case where one of the two springpots is reduced to an elastic spring ($\beta=0$). In this case the rheological constitutive equation reads:

%+++++++++++++++++++++++++
\begin{equation}
	\frac{d^\alpha \gamma}{dt^\alpha} = \frac{1}{G_0}\left[\frac{\sigma(t)}{\tau_{FM}^\alpha}+\frac{d^\alpha \sigma}{dt^\alpha}\right]
	\label{eqn:FMM_constitutive}
\end{equation}
%+++++++++++++++++++++++++

\noindent
Note that, for $\alpha=1$, Eq.~\ref{eqn:FMM_constitutive} corresponds to a standard Maxwell fluid.

	Solutions of the FM model for standard rheological experiments have been previously computed \cite{bagley_power_1989, jaishankar_power-law_2013, jaishankar_fractional_2014, jozwiak_fractional_2015}.
	For a sinusoidal deformation of angular frequency $\omega$, the storage, $G'$, and loss, $G"$, moduli read:

%+++++++++++++++++++++++++
\begin{equation}
	G'(\omega) =   G_0 \zeta \frac{\cos\left(\frac{\pi}{2}\alpha\right) + \zeta}{1+\zeta^2+2\zeta \cos\left(\frac{\pi}{2}\alpha\right)}
	\label{eqn:fmm_oscill_storage}
\end{equation}
\begin{equation}
	G''(\omega) =   G_0 \zeta \frac{\sin\left(\frac{\pi}{2}\alpha\right)}{1+\zeta^2+2\zeta \cos\left(\frac{\pi}{2}\alpha\right)}
	\label{eqn:fmm_oscill_loss}
\end{equation}
%+++++++++++++++++++++++++

\noindent
where $\zeta=( \omega\tau_{\rm{FM}})^\alpha$.

	On the other hand, the creep deformation following the application at time $t=0$ of a step stress of amplitude $\sigma_0$ reads:

%+++++++++++++++++++++++++
\begin{equation}
	\gamma(t)=\frac{\sigma_0}{G_0} \left[1 + \frac{1}{\Gamma(\alpha+1)}\left(\frac{t}{\tau_{\rm{FM}}}\right)^\alpha\right]
	\label{eqn:FMM_creep}
\end{equation}
%+++++++++++++++++++++++++

\noindent
This equation can be decomposed as follows:

%+++++++++++++++++++++++++
\begin{equation}
	\gamma(t)=\gamma_e^+ + \gamma_{\rm{creep}}(t)
	\label{eqn:FMM_creep2}
\end{equation}
%+++++++++++++++++++++++++

\noindent
Here $\gamma_e^+=\sigma_0/G_0$ is the instantaneous elastic part of the mechanical response and $\gamma_{\rm{creep}}(t)=\frac{\gamma_e}{\Gamma(\alpha+1)}\left(\frac{t}{\tau_{\rm{FM}}}\right)^\alpha$ is analogous to Eq.~\ref{eqn:springpot_creep} and corresponds to the cumulated creep deformation since the application of the step stress at time $t=0$.

	It is also interesting to investigate the creep recovery, that is the time evolution of the deformation following the release of the stress $\sigma_0$ at time $T$. We here define $t'=t-T$ as the time elapsed since the release of the stress, which has been applied from time $t=0$ to time $T$. Similarly to the creep, the creep recovery is composed of an instantaneous elastic relaxation of amplitude $\gamma_e^-=\sigma_0/G_0$ and a slow decay function $\gamma_{\rm{rec}}(t')$ that reads:

%+++++++++++++++++++++++++
\begin{equation}
\gamma_{\rm{rec}}(t')=\gamma_{\rm{creep}}(T)\left[\left(1+\frac{t'}{T}\right)^\alpha-\left(\frac{t'}{T}\right)^\alpha\right]
\label{eqn:FMM_recovery}
\end{equation}
%+++++++++++++++++++++++++

	We observe that the linear viscoelastic creep of a FM model is completely reversible, since $\gamma_{\rm{rec}}(t'\rightarrow\infty)=0$. Interestingly, Eq.~\ref{eqn:FMM_recovery} predicts that the initial un-deformed configuration is recovered with a characteristic time that uniquely depends on the duration $T$ of the creep, whereas it is independent of the natural timescale $\tau_{\rm{FM}}$.
	We show in Fig.~\ref{fig:springpot}b the creep and recovery for a Fractional Maxwell model with a representative value of $\alpha=0.5$ together with the asymptotic behaviors, a Maxwell fluid ($\alpha=1$) and an elastic solid ($\alpha=0$).

\section{Material and methods}
\label{sec:matmeth}

\subsection{Sample}

	We investigate a fractal colloidal gel. The gel is formed by aggregating a water suspension of charged silica particles (Ludox TM50, from Sigma Aldrich) {\color {myc} (of diameter $25$ nm)} at a volume fraction $\phi=5\%$. Particle aggregation is triggered by increasing \textit{in situ} the ionic strength of the solvent, thanks to the hydrolysis of urea into carbon dioxide and ammonia: $CO(NH_2)_2+H_2O \rightarrow CO_2 + 2NH_3 \rightarrow NH_4^+ +NH_3 + HCO_3^-$. This chemical reaction is catalyzed by an enzyme (Urease U1500-20KU, from Sigma Aldrich), which increases the ionic strength of the solvent and thus screens the electrostatic repulsion between particles, eventually inducing particle aggregation \cite{wyss_small-angle_2004}. Adding {\color {myc} urea (concentration $1$ Mol/l) and urease ($35$ U/ml) to the colloidal suspension at room temperature induces a sol-gel} transition of the suspension within roughly $3$ hours.
	Structural information on the gel and on the particles are obtained by neutron and light scattering techniques. In order to increase the contrast between the particles and the solvent in neutron scattering, water ($\rm{H}_2\rm{O}$) is replaced by heavy water ($\rm{D}_2\rm{O}$), without
any indication of any structural and rheological alteration of the sample following the replacement of $\rm{H}_2\rm{O}$ by $\rm{D}_2\rm{O}$.

\subsection{Scattering techniques and rheology}

	The structure of the sample is probed combining different scattering techniques, allowing one to access a wide range of scattering vectors $q$, from $0.1$ to $2000$ $\mu\rm{m}^{-1}$. We use small-angle neutron scattering (SANS) (PA20 beamline at Laboratoire L\'eon Brillouin, France), and custom made wide-angle (WALS) and small-angle light scattering setups (SALS) \cite{tamborini_multiangle_2012}.
For SANS, the sample is held in a $2$ mm thick rectangular cell, whereas for light scattering experiments the cell thickness was decreased to 1 mm in order to avoid multiple scattering.

	Rheological measurements are performed in the Couette cell of a stress-controlled rheometer (Anton Paar MCR502), with a low viscosity silicon oil film on top of the sample to prevent evaporation over very long timescales (several weeks). Frequency sweep, respectively creep, measurements are performed at sufficiently small strain amplitude (typically $\gamma_0=0.1$ \%), resp. at sufficiently small applied stress, to ensure data are acquired in the linear regime. Strain sweep experiments are also conducted at $1$ Hz to determine the upper bound of the linear regime. {\color {myc} For the frequency sweep during sample aging, we choose a protocol that minimizes the impact of the finite experimental time on the measured viscoelastic moduli: we performe high-to-low and low-to-high frequency sweeps one after the other, and for each frequency we take the average of the moduli measured during the two sweeps. The waiting time $t_w$ reported in each plot corresponds to the average between the times of the two measurements, and is the same for all frequencies. Except at very short aging time, the duration of one frequency sweep is short compared to the sample aging time, hence ensuring reliable results.}

	In addition, the sample structure, dynamics and rheological properties are also probed simultaneously using a custom made SALS apparatus~\cite{tamborini_multiangle_2012}, which is coupled to a linear, parallel plate stress-controlled shear cell~\cite{aime_stress-controlled_2016}. The setup allows one to access scattering vectors $q$ ranging from about $0.2$ $\um^{-1}$ to $4$ $\um^{-1}$, essentially oriented in the (vorticity, velocity) plane, with a minor $q$-dependent component along the shear gradient direction.
	Static and dynamic light scattering experiments under oscillatory shear are performed as described in \cite{aime_microscopic_2018-1}. In static light scattering experiments one measures the time-averaged intensity scattered as a function of the scattering vector $q$, from which information on the particle size and interparticle correlations can be extracted \cite{zemb_neutron_2002}. On the other hand, in dynamic experiments one measures the two-time intensity correlation function $g_2(\vec{q}, t_1, t_2) - 1 \propto \left|g_1(\vec{q}, t_1, t_2)\right|^2$, where $g_1$ is the intermediate scattering function, quantifying the relative displacements of the scatterers along the scattering vector $\vec{q}$~\cite{berne_dynamic_2013}.
	Under shear, the field describing particle displacements is generally anisotropic, and can be decoupled into an affine contribution, always in the shear direction $\hat{e}_\parallel$, and a nonaffine contribution, which typically has nonzero components perpendicular to $\hat{e}_\parallel$. In the scattering geometry, it is thus possible to decouple the two contributions by looking at scattering vectors oriented in the vorticity direction ($\vec{q}_\perp$), which only carry information about nonaffine rearrangements whereas scattering vectors oriented in the velocity direction ($\vec{q}_\parallel$) contain contributions from both affine and non-affine dispalcements.

	All experiments are performed at room temperature. For both scattering measurements and rheological measurements, the sample is loaded in the scattering or shear cell while still in a liquid state and is then allowed to gel \textit{in-situ}. By monitoring the sample at rest over long time, we observe that the gel is stable, and we can neglect the influence of gravity on the microscopic structure.

\section{Results}
\label{sec:experimental}

\subsection{Sample structure}

%------------------------------------------------
\begin{figure}[h]
\centering
  \includegraphics[width=0.5\columnwidth]{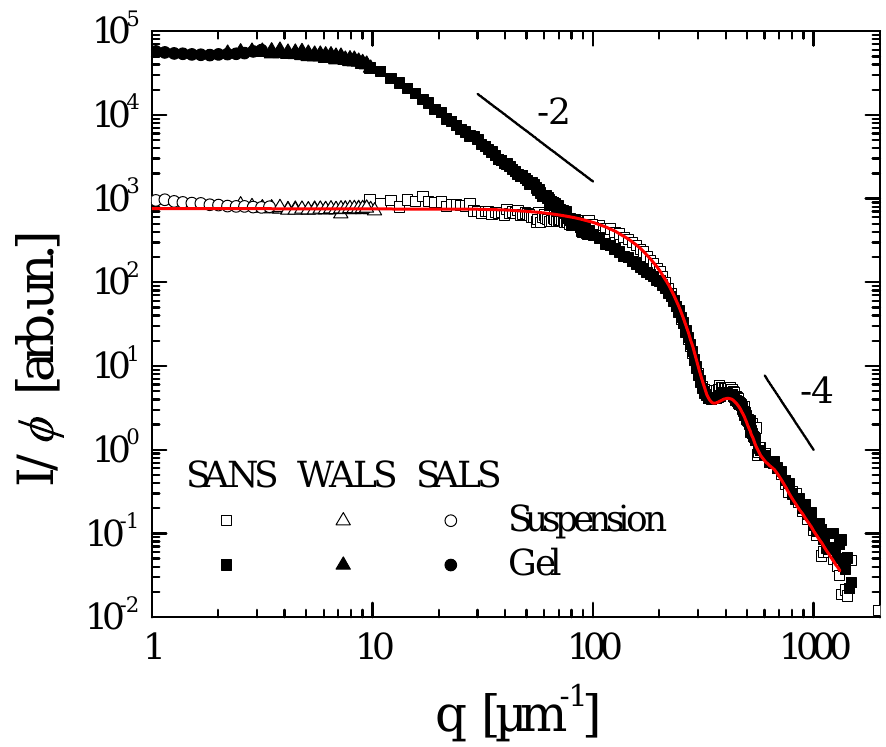}
  \caption{Scattered intensity of the colloidal gel at $\phi=5$\% volume fraction (filled symbols) and of a stable colloidal suspension, at $\phi=0.037$\% (empty symbols), probed with small angle neutron scattering, wide-and X ray scattering and static light scattering, as indicated in the legend. Red line: fitted form factor of a polydisperse set of spheres of average diameter $25$ nm and polydispersity $10\%$.}
  \label{fig:Iofq}
\end{figure}
%------------------------------------------------

	We show in Fig.~\ref{fig:Iofq} the scattering profiles, i.e. the scattered intensity, $I$, normalized by particle volume fraction, $\phi$, of a dilute, liquid and stable suspension ($\phi=0.037$ \%) and of a colloidal gel ($\phi=5$ \%) as a function of the scattering vector $q$, over several orders of magnitude, thanks to the combination of three experimental techniques (SANS, WALS and SALS). Over the whole range of scattering vectors investigated, the scattering profile of the dilute suspension can be quantitatively accounted for by the form factor of polydisperse spherical particles \cite{bresler_sasfit_2015}: $I(q)=\int \rho(R) P(q,R) dR$, where $P(q,R)= \left[3\left(\sin(qR)-qR\cos(qR)\right)\left(qR\right)^{-3}\right]^2$ is the form factor of a uniform sphere of radius $R$, and the weighting function $\rho(R)$ is derived from a Gaussian distribution of particle sizes. The best fit of the experimental data (continuous line in Fig.~\ref{fig:Iofq}) yields an average diameter $a=25$ nm and a $10$ \% polydispersity (defined as the full width at half maximum of the particle size distribution).
	For the colloidal gel, one sees that the data at large $q$, which probe the structure of the individual particles, perfectly superimpose with the measurement for the dilute suspension, as expected. By contrast, data for $q<200$ $\mu\rm{m}^{-1}$ largely differ. The scattered intensity is the one expected for crowded fractal clusters of particles~\cite{carpineti_transition_1993, manley_gravitational_2005}. For $q$ in the range $(10-200)$ $\mu\rm{m}^{-1}$, $I$ decays as $q^{-2}$, indicating a fractal dimension $d_f=2$.
	At very low $q$ ($q<q^*$, with $q^*\approx10$ $\mu\rm{m}^{-1}$), the scattered intensity becomes almost $q$-independent, indicating that on length scales larger than $2 \pi/q^* \approx 0.6$ $\mu\rm{m}$, the sample structure is rather homogeneous. This value is in good numerical agreement with the theoretical expectation for the cluster size  $\xi \sim a \phi^{1/(d_f-3)} \sim 1$ $\mu\rm{m}$ \cite{carpineti_spinodal-type_1992}.

\subsection{Gelation dynamics}

\begin{figure}[h]
\centering
  \includegraphics[width=0.5\columnwidth]{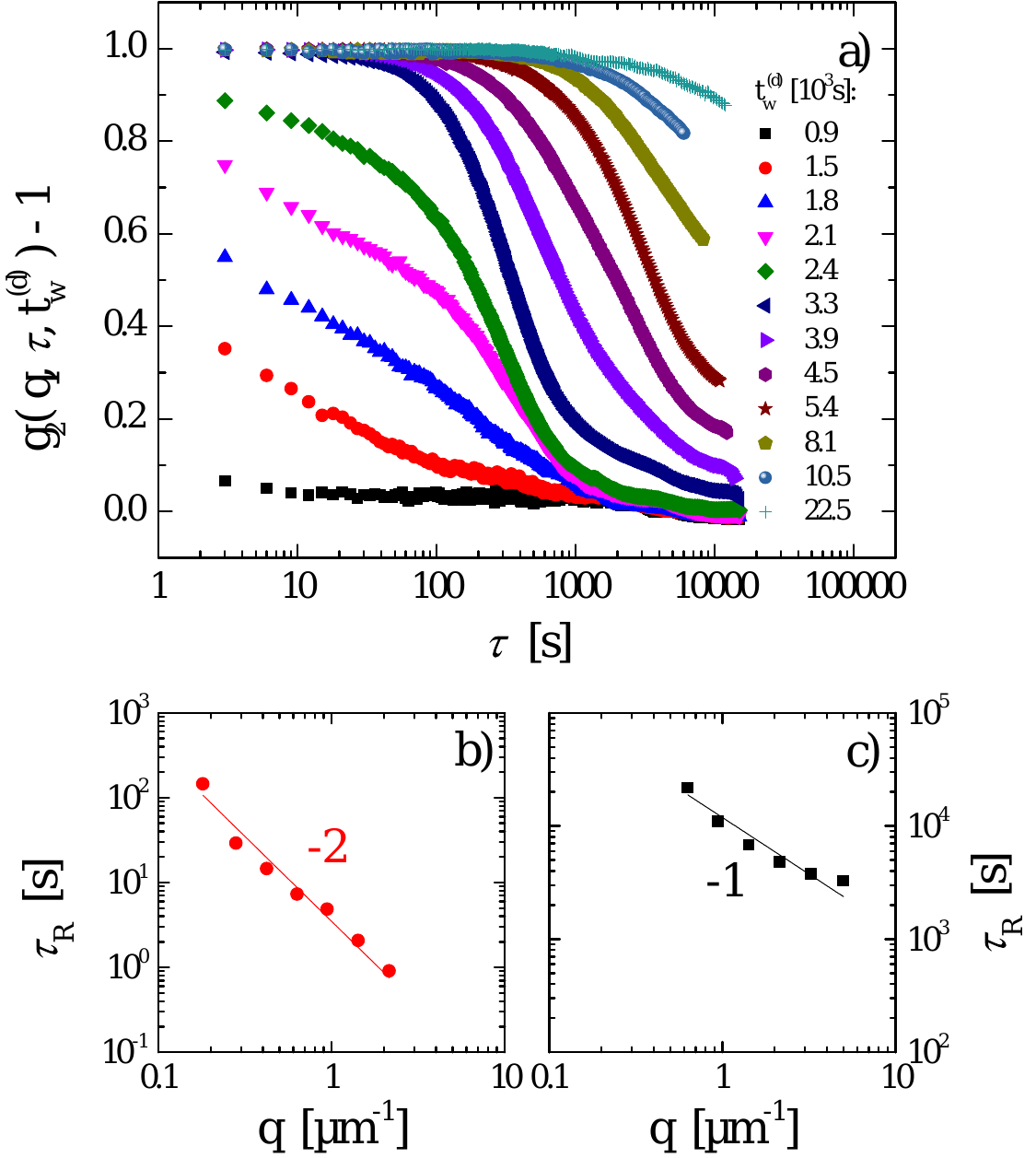}
  \caption{(a) Intensity correlation functions plotted as a function of time delay, $\tau$, for different times $t_w^{(d)}$ after the onset of aggregation. The scattering vector is $q=1.5$ $\mu\rm{m}^{-1}$. {\color {myc} (b,c)} Relaxation times  extracted from correlation functions during the aggregation stages ($t_w^{(d)}=1200$~s, red circles) and once a network is formed ($t_w^{(d)}=5000$~s, black squares), and plotted as a function of $q$. The lines are power law fits with exponent $-2$ (aggregation stage) {\color {myc} (b)} and $-1$ (gel stage) {\color {myc} (c)}.}
  \label{fig:GelDynamics}
\end{figure}

	Gelation dynamics is probed by dynamic light scattering. At early stages, right after preparation, the sample is a liquid with a viscosity comparable to that of water. As a consequence of enzymatic activity, the ionic strength of the solvent grows in time, and after an induction time that depends on the amount of enzyme added (typically $8000$ s in our experiments), it becomes large enough to trigger particle aggregation. This event can be clearly seen with dynamic light scattering, since the formation of larger particle clusters slows down the microscopic dynamics.
	A few minutes after the onset of aggregation, the relaxation time $\tau_R$ becomes measurable for most accessed scattering vectors. The scaling $\tau_R \sim q^{-2}$ confirms that we probe the diffusive motion of the aggregates. With time, the decay of the correlation functions shifts towards larger delays, showing that the aggregates are growing in size. Eventually, clusters become large enough to touch each other, and at that moment a system-spanning network is formed. When this happens, the dynamics slows down tremendously and changes qualitatively, with correlation functions changing from simple exponentials to compressed exponentials and the $\tau_R$ scaling changing from $q^{-2}$ (diffusive-like dynamics) to $q^{-1}$ (ballistic-like dynamics) (Fig.~\ref{fig:GelDynamics}c), as observed for many soft solid materials \cite{cipelletti_universal_2000, cipelletti_slow_2005}.

%------------------------------------------------
%------------------------------------------------

\subsection{Linear viscoelasticity}

\subsubsection{Frequency sweep during aging}

\begin{figure}[h]
\centering
  \includegraphics[width=0.5\columnwidth]{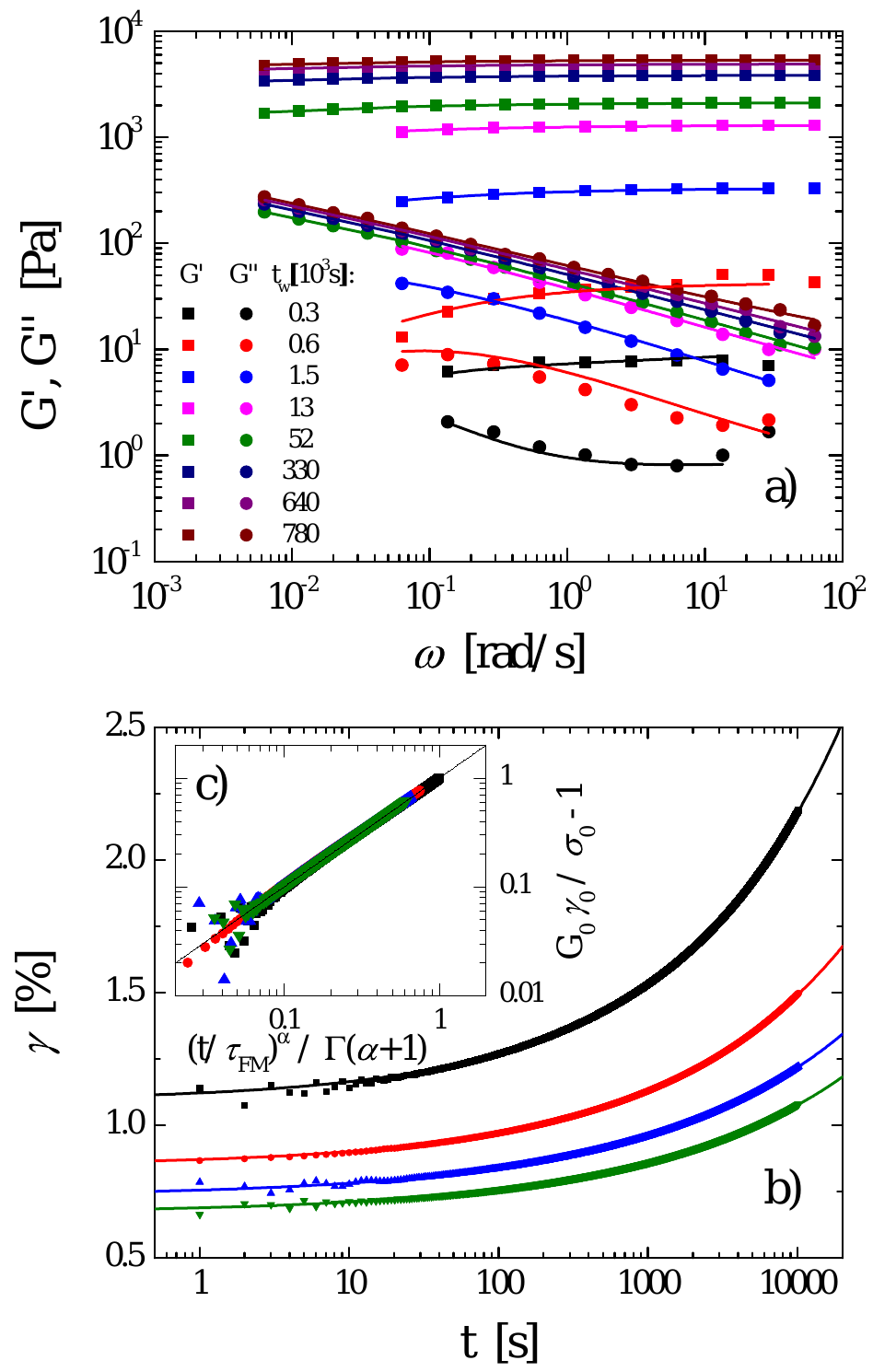}
  \caption{(a) Viscoelastic moduli measured during sample aging at a strain amplitude $\gamma_0 = 0.1\%$, and plotted as a function of angular frequency for different waiting times $t_w$ as indicated in the legend. Solid lines represent fits using Eqs.~\ref{eqn:fmm_oscill_storage} and \ref{eqn:fmm_oscill_loss}. The fit parameters (stars) are displayed in figure~\ref{fig:fmm_fitparams}. (b) Creep deformation under a shear stress $\sigma_0=30$~Pa for 4 different sample ages, {\color {myc} from top to bottom $t_w= 12\times10^4, 26\times10^4, 40\times10^4, 54\times10^4 s$}. Lines are fits of the experimental data using Eq.~\ref{eqn:FMM_creep}. The fit parameters (filled squares) are shown in Fig.~\ref{fig:fmm_fitparams}.  {\color {myc} (c) Same data as in (b) plotted in rescaled units (see text) (symbols) and theoretical expectation (line).} }
  \label{fig:aging_dfs}
\end{figure}

	As a network forms, a finite storage modulus $G'$ is measured by rheometry. Subsequently, the rheological response of the material rapidly evolves with time  towards the elasticity-dominated pattern shown in Fig.~\ref{fig:aging_dfs}a. One defines the sample age $t_w$ as the time elapsed since the gelation time, chosen at the time of crossover between $G'$ and $G''$, as measured at an angular frequency $\omega=0.628$ rad/s.

%------------------------------------------------
%------------------------------------------------
	
	We probe the evolution with sample age of the frequency dependence of the complex modulus. Data are measured for $t_w$ spanning more than $3$ orders of magnitude (from $300$ s to $8\times10^5$ s). Over this timescale, the storage modulus increases by almost $3$ orders of magnitude, and the loss modulus by more than $2$ orders of magnitude. Interestingly, all along the aging process, the sample viscoelasticity is always very well described by the Fractional Maxwell model~\cite{jaishankar_power-law_2013}, as shown in Fig.~\ref{fig:aging_dfs}a where the best fits of the experimental data using Eqs.~\ref{eqn:fmm_oscill_storage} and \ref{eqn:fmm_oscill_loss} are displayed.
	
\begin{figure}[h]
\centering
  \includegraphics[width=0.5\columnwidth]{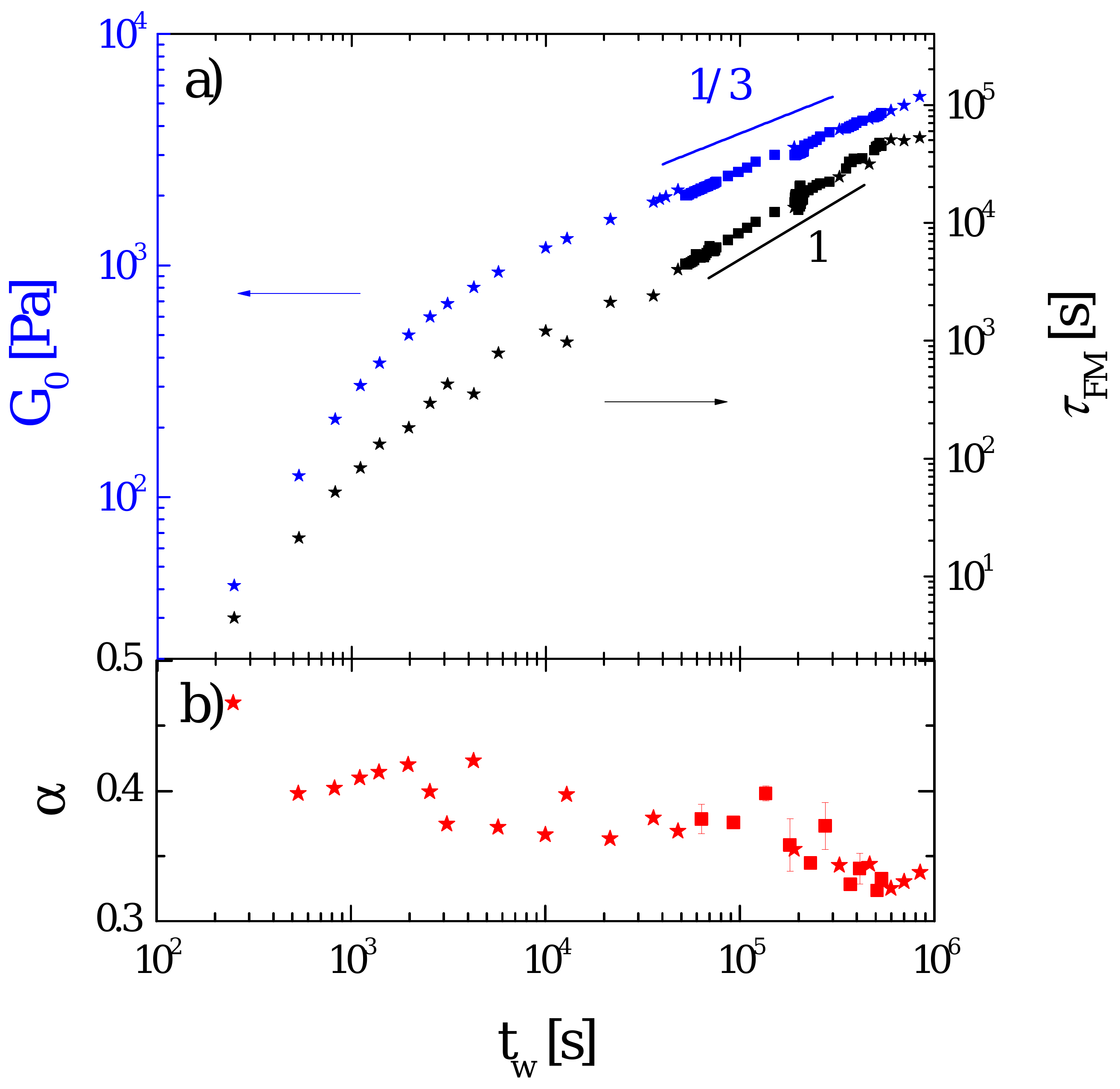}
  \caption{Fractional Maxwell parameters extracted from both frequency sweeps (stars) and creep data (squares) fitted by Eqs.~\ref{eqn:fmm_oscill_storage}, \ref{eqn:fmm_oscill_loss} and \ref{eqn:FMM_creep} respectively.}
  \label{fig:fmm_fitparams}
\end{figure}

	The evolution of the fit parameters $\alpha$, $\tau_{\rm{FM}}$ and $G_0$ upon sample aging are given in Fig.~\ref{fig:fmm_fitparams}. Although the data are somehow noisy, the exponent $\alpha$ is measured to slightly decrease with sample age, from $\sim 0.42$ to $\sim 0.32$, in analogy to what found in the literature on biopolymer gels \cite{curtis_validation_2015}. On the other hand, the elastic modulus, $G_0$, and the characteristic time, $\tau_{\rm{FM}}$, display a smooth and continuous increase with $t_w$. Both parameters are measured to increase rather rapidly at short time ($t_w\lesssim 10^4$ s), and less steeply at longer times. In the late time regime, we find $\tau_{\rm{FM}}\sim t_w$, with $\tau_{\rm{FM}}$ about one order of magnitude smaller than the waiting time and $G_0 \sim t_w^{1/3}$. {\color {myc} Note that the scaling law measured for the elastic modulus with sample age is similar to previous findings for other colloidal gels \cite{manley_time-dependent_2005, koumakis_two_2011-1, guo_gel_2011-1, calzolari_interplay_2017}}.

%------------------------------------------------
%------------------------------------------------

\subsubsection{Creep experiments for old samples}

\begin{figure}[h]
\centering
  \includegraphics[width=0.5\columnwidth,clip]{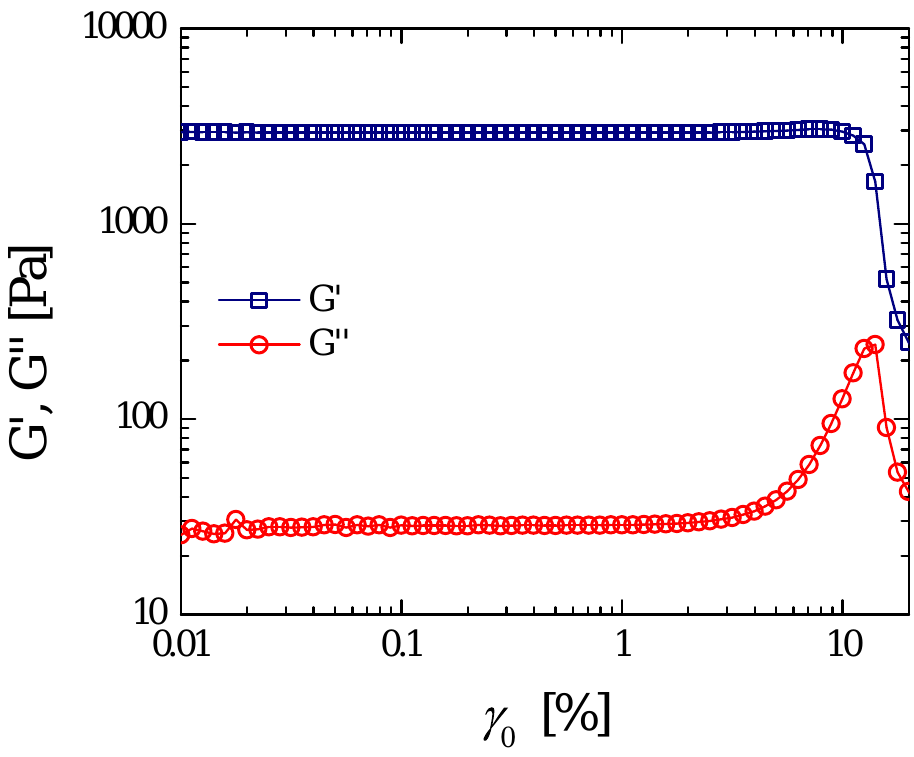}
  \caption{Viscoelastic moduli $G'$ (squares) and $G''$ (circles) probed at angular frequency $\omega=6.28$ Hz {\color {myc} as a function of the strain amplitude, $\gamma_0$. $\gamma_0$ is increased from $0.01$\% to $1000$\% at a rate of $20$ points per decade. $10$ oscillations are performed for each $\gamma_0$. Sample age is $240000$ s.}}
  \label{fig:laos}
\end{figure}

To check for consistency as well as to probe smaller frequencies, we perform creep experiments in the linear regime. Typical creep profiles are shown in Fig.~\ref{fig:aging_dfs}b for different waiting times $t_w$, in the range $(1.2-5.4)\cdot 10^5$ s. Here the sample is rather old, ensuring that during each creep measurement, which lasts $10000$ s at most, sample aging is negligible. The shear stress is fixed at $\sigma_0=30$ Pa and is about 2 orders of magnitude smaller than the elastic modulus, such that measurements are performed in the linear regime. Indeed, a strain sweep on an old sample ($t_w=24 \times 10^5$ s) shows that linearity extends beyond $1\%$ deformation (Fig.~\ref{fig:laos}).

%------------------------------------------------
%------------------------------------------------

	As the sample ages and $G_0$ increases, the initial elastic jump for a fixed applied stress becomes smaller, and the further increase with time of the strain becomes smaller as well. For all sample ages investigated (between $10^5$ and $10^6$ s), we find that the whole time evolution of the strain is very well fitted by the theoretical prediction of the FM model (Eq.~\ref{eqn:FMM_creep}), with parameter values that are fully consistent with those measured in oscillatory experiments (see squares in Fig. \ref{fig:fmm_fitparams}). {\color {myc} To further prove the appropriateness of the FM model, we plot the data in rescaled units, $\tilde{X}=\frac{1}{\Gamma(\alpha+1)} \left(\frac{t}{\tau_{\rm{FM}}}\right)^{\alpha}$ and $\tilde{Y}= \frac{G_0 \gamma}{\sigma}-1 $. We find a nice collapse onto a single curve, which is moreover in perfect agreement with the theoretical expectation $\tilde{Y}= \tilde{X}$ (Eq.~\ref{eqn:FMM_creep}).}

%------------------------------------------------
\begin{figure}[h]
\centering
  \includegraphics[width=0.5\columnwidth]{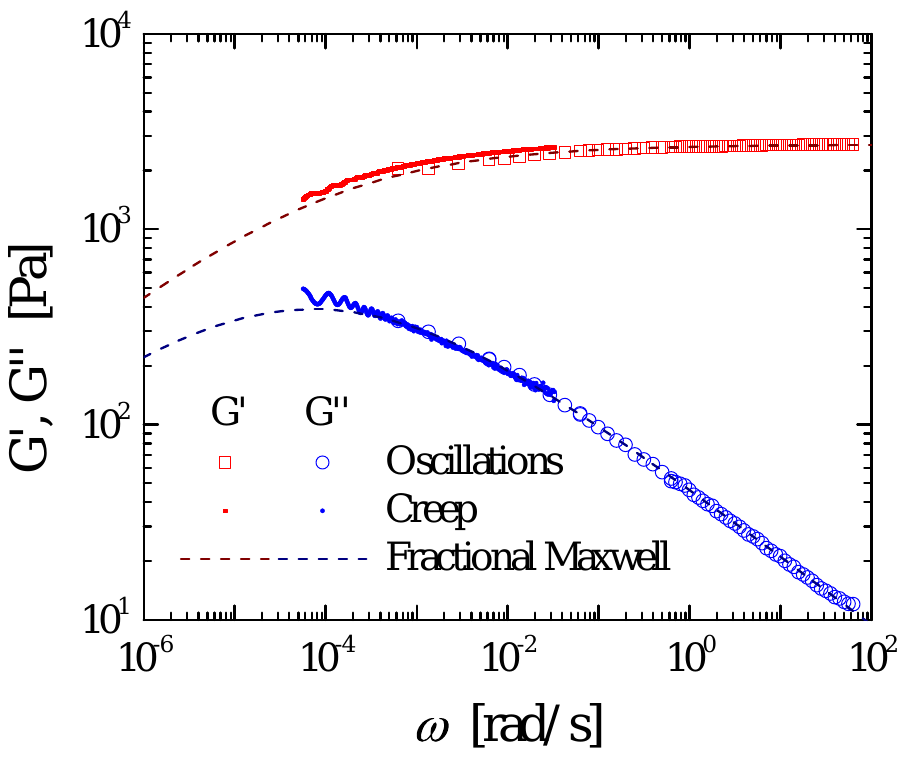}
  \caption{Comparison between viscoelastic moduli probed in small amplitude oscillatory strain at $\gamma_0=0.1\%$ amplitude (large open symbols) and creep under a constant shear stress $\sigma_0/G_0=1.8\%$ (small solid points) for a sample age $t_w = 2 \cdot 10^5$ s. Dashed lines represent best fits using the Fractional Maxwell model. The fit parameters are: $G_0=2700$~Pa, $\tau_M=1.5\cdot10^4$~s, $\alpha=0.35$.}
  \label{fig:dfs_wide}
\end{figure}
%------------------------------------------------

	A direct comparison of creep and frequency sweep measurements can be achieved using methods available in literature~\cite{evans_direct_2009}.  As an illustration, we show in Fig.~\ref{fig:dfs_wide} the direct measurement of the frequency-dependence of the storage and loss moduli, together with the frequency dependence of $G'$ and $G"$ computed from a creep experiment \cite{evans_direct_2009}, and the fit using the FM model. We find a nice overlap of the two sets of data. Importantly, creep measurements allow the measurements to be extended at lower frequency as compared to oscillatory rheology, about one decade for the data shown in Fig.~\ref{fig:dfs_wide}. These data allow one to probe the sample behavior almost up to the FM characteristic time $\tau_{FM}$. At low angular frequencies ($\omega \ll \tau_{\rm{FM}}^{-1}$), both $G'$ and $G''$ increase as $\omega^\alpha$, whereas in the high angular frequency regime ($\omega \gg \tau_{\rm{FM}}^{-1}$), $G'\sim G_0$ and $G'' \propto \omega^{-\alpha}$.
	
	Overall, we find that the evolution of the three parameters of the FM model as extracted from oscillatory shear and creep experiments nicely overlap over the whole range of sample ages. This demonstrates the relevance of the Fractional Maxwell model to describe the linear viscoelasticity of our fractal colloidal gel.

\subsubsection{Creep-recovery experiments}

\begin{figure}[h]
	\centering
	\includegraphics[width=0.5\columnwidth]{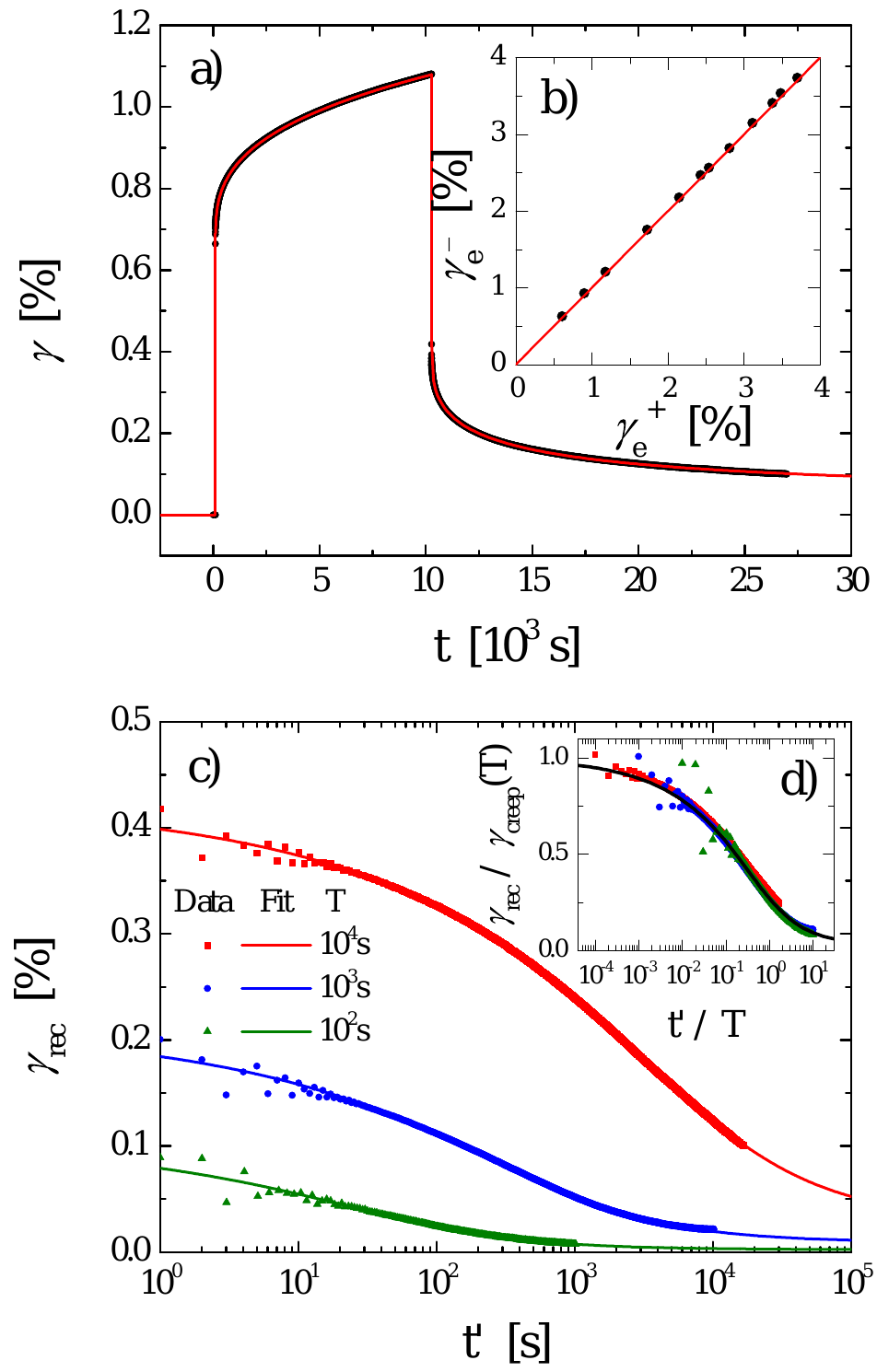}
	\caption{(a) Viscoelastic creep under a constant shear stress $\sigma_0=30$~Pa, lasting a time $T=10000$~s and followed by creep recovery. The symbols are experimental data and the line is a fit with the FM model yielding $G_0=4400$~Pa, $\tau_{FM}=4.8\cdot10^4$~s, $\alpha=0.33$. (b) Recovery downwards jump $\gamma_e^-$ plotted as a function of the instantaneous upward elastic jump $\gamma_e^+=\sigma_0/G_0$. Red line represents the expected $\gamma_e^+=\gamma_e^-$ behavior. (c) Creep recovery curves following a creep under a stress $\sigma_0=30$~Pa lasting $T$ seconds, as specified in the legend. The symbols are the experimental data points and the line is the best fit with FM model. (d) Same data plotted in rescaled units,  $\gamma_{\rm{rec}}/\gamma_{\rm{creep}}$ as a function of the rescaled time $t'/T$, as in Eq. \ref{eqn:FMM_recovery}. The sample age is $t_w\approx 5 \cdot 10^5$~s.}
	\label{fig:creep_and_recovery}
\end{figure}

	To further test the FM model, we perform creep recovery experiments. Here, after a creep of duration $T$, the stress is released, and the sample is left at rest until the creep strain is fully relaxed, before applying another step stress. $T$ is chosen such that the sample age ($t_w>10^5$ s) is much larger than the duration of the experiment, in order to keep aging effects negligible during creep. The strain evolution during a representative creep and recovery experiment is shown in Fig.~\ref{fig:creep_and_recovery}a, together with the fit with the FM model (Eq.~\ref{eqn:FMM_creep2} for the creep and Eq.~\ref{eqn:FMM_recovery} for the recovery).
	We find that the FM reproduces extremely well the experimental data. Interestingly, we measure that, for initial elastic jumps $\gamma_e^+$ up to $3.5\%$ (obtained by imposing various stress amplitudes), the instantaneous recovery $\gamma_e^-$ is exactly equal to $\gamma_e^+$ (inset Fig.~\ref{fig:creep_and_recovery}). Such equality provides a further indication that the experimental data correspond to the linear viscoelastic regime, in agreement with the findings for the oscillatory and creep experiments.

%------------------------------------------------
%------------------------------------------------

	Typical strain relaxations during creep recovery are shown in Fig.~\ref{fig:creep_and_recovery}b for an old sample ($t_w\approx 5 \cdot 10^5$s) that has been submitted to three consecutive creep tests of duration $T=100$, $1000$ and $10000$~s. We find that as the duration of the creep is longer, the initial value of $\gamma$ in the creep recovery is higher, as expected.
	Equation~\ref{eqn:FMM_recovery} predicts that after the sudden release of the external stress the macroscopic deformation should decay to $0$ with a characteristic shape depending on $\alpha$ and scaling as $t'/T$. Figure~\ref{fig:creep_and_recovery}d shows that the $t'/T$ scaling does indeed hold, since data measured for different creep durations collapse on a single curve once the strain is normalized by the strain cumulated at the end of the previous creep step ($\gamma_{\rm{creep}}(T)$, cf. Eq.~\ref{eqn:FMM_creep2}), and time $t'$ is normalized by the creep duration $T$.

	The data shown in Fig.~\ref{fig:creep_and_recovery} show that creep recovery experiments are also in good quantitative {\color {myc} agreement} with FM predictions, and that the cumulated strain $\gamma$ almost fully relaxes to $0$. This suggests that the rearrangements occurring during creep are reversible, despite the sample is not purely elastic.

\subsection{Reversibility}

\subsubsection{Structural anisotropy}

%------------------------------------------------
\begin{figure}[h]
\centering
  \includegraphics[width=0.5\columnwidth,clip]{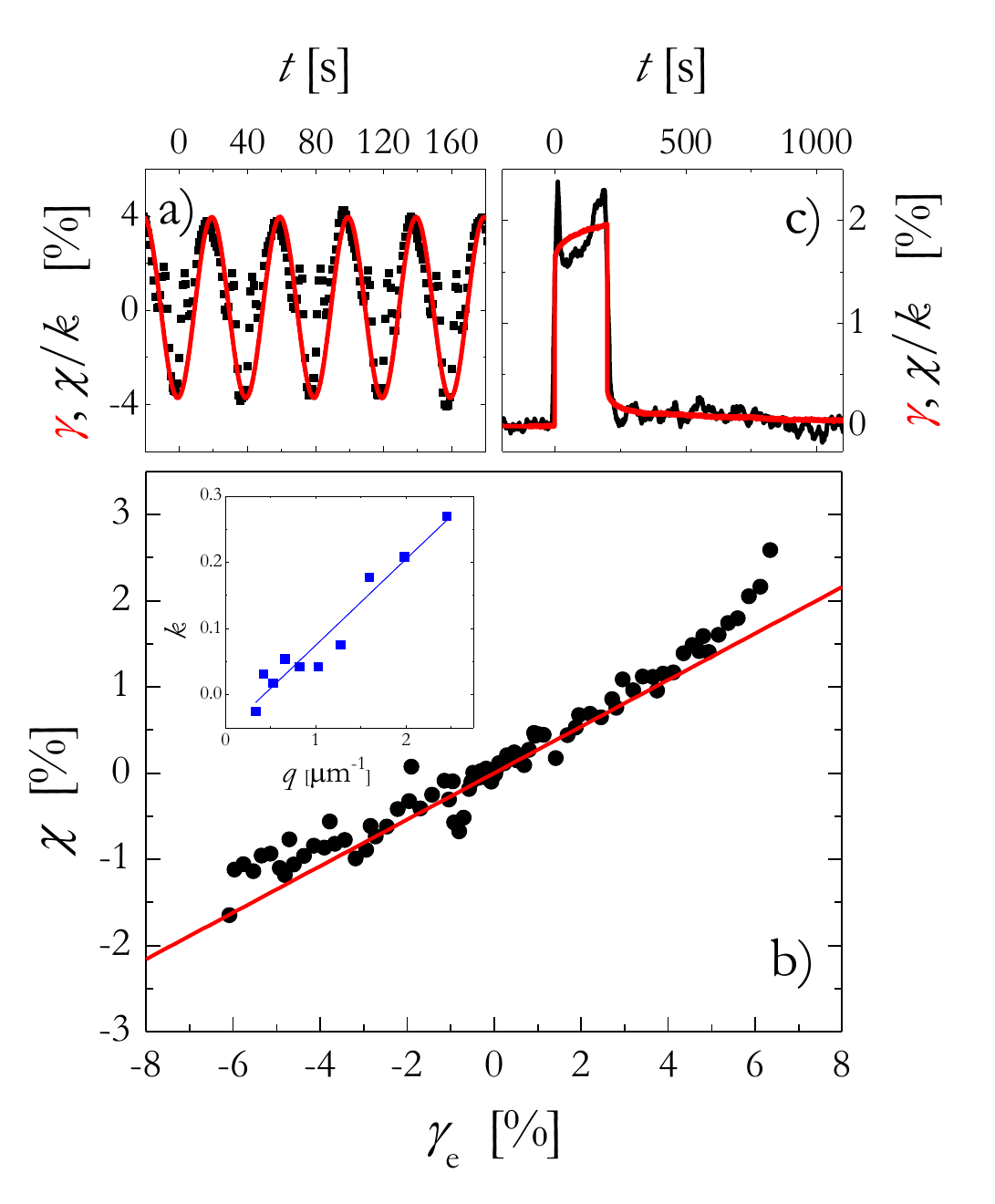}
  \caption{a) {\color {myc} Red} symbols: oscillating strain under an oscillating stress of amplitude $\sigma_0/G_0={\color {myc} 4}\%$. {\color {myc} Black} line: static asymmetry $\chi(q)$ measured at $q=2.6\mu\textrm{m}^{-1}$ divided by the proportionality constant $k(q)$ discussed in the text. b) Symbols: static asymmetry $\chi(q=2.6\mu\textrm{m}^{-1})$ observed under a step strain deformation of amplitude $\gamma_e$. Positive and negative $\gamma_e$ values represent step strains in opposite directions. Red line: linear fit of the small deformation regime, used to extract the proportionality constant $k(q)$. {\color {myc}Inset: $q$ dependence of the proportionality constant. The line is a linear fit of the experimental data (symbols).} c) {\color {myc} Red} curve: creep deformation ($\sigma_0/G_0=1.5\%$) and recovery after a creep time $T=200$~s. {\color {myc} Black} curve: static asymmetry $\chi(q=2.6\mu\textrm{m}^{-1})$ divided by $k(q)$. {\color {myc} Sample age is $t_w=1.5\times10^5$ s.}}
  \label{fig:asym}
\end{figure}
%------------------------------------------------

	We check for reversibility at a microscopic level by direct structural measurements during a sinusoidal shear deformation, using the home-made set-up that couples small-angle light scattering to a stress-controlled shear cell described in~\cite{aime_stress-controlled_2016}.
	The microscopic structure of the sample is monitored as a function of time during sample deformation for rheological tests similar to those performed in a classical rheometer and presented above. Over the range of wave-vectors $q$ probed (between $0.4$ and $4$ $\um^{-1}$), the scattered intensity does not evolve, suggesting that the structure of the colloidal gel is fundamentally preserved under deformation in the linear regime. However, as the sample is sheared, a small anisotropy (a few percents) in the static structure factor is detected. Such anisotropy has already been observed in simulations {\color {myc} \cite{park_structure-rheology_2017, colombo_stress_2014, bouzid_network_2018, jamali_microstructural_2017}} and measured in experimental works on colloidal gels \cite{mohraz_orientation_2005, vermant_flow-induced_2005, kim_microstructure_2014}. It has been interpreted as the signature of structural alignment in the direction of flow.
	
	In particular, in our experiments we find that the intensity of the light scattered in the direction perpendicular to the shear direction (vorticity direction), $I(\vec{q}_\perp)$, is constant, whereas the one scattered in the parallel (velocity) direction, $I(\vec{q}_\parallel)$, changes roughly linearly with the macroscopic deformation. To quantify this effect, we define a static anisotropy parameter $\chi(q)=\left[I(\vec{q}_\parallel) - I(\vec{q}_\perp)\right]/\left[I(\vec{q}_\parallel) + I(\vec{q}_\perp)\right]$.  We find that the anisotropy $\chi(q)$ nicely follows the evolution of the strain when the sample is submitted to a small oscillatory stress, as illustrated in Fig.~\ref{fig:asym}a for $\sigma_0/G_0=4\%$ and $q=2.6 \um^{-1}$. Similarly, $\chi$ {\color {myc} roughly} follows the time evolution of the measured macroscopic strain during a creep and creep recovery experiment (Fig.~\ref{fig:asym}c). Overall, our data show that, in the linear regime, the structural anisotropy is proportional to the strain applied and is therefore fully reversible.
	More quantitatively, we plot $\chi$ as a function of the strain amplitude $\gamma_e$ in Fig.~\ref{fig:asym}b. For $-4 \% < \gamma_e < +4 \%$, we can empirically model the observed asymmetry with the linear relation $\chi(q, \gamma)=k(q)\gamma$, with the factor $k(q)$ a phenomenological proportionality constant that increases roughly linearly with $q$ {\color {myc} (inset of Fig.~\ref{fig:asym}b)}. Consistently, $4\%$ correspond to the limit of the linear regime as defined from the maximum strain amplitude at which the complex moduli are independent of the strain amplitude (Fig.~\ref{fig:laos}). Beyond $\gamma = 4\%$, an excess of anisotropy is measured, corresponding to the onset of rheological non-linearity.

\subsubsection{Non-affine dynamics and microscopic reversibility}

\begin{figure}[h]
  \includegraphics[width=0.5\columnwidth]{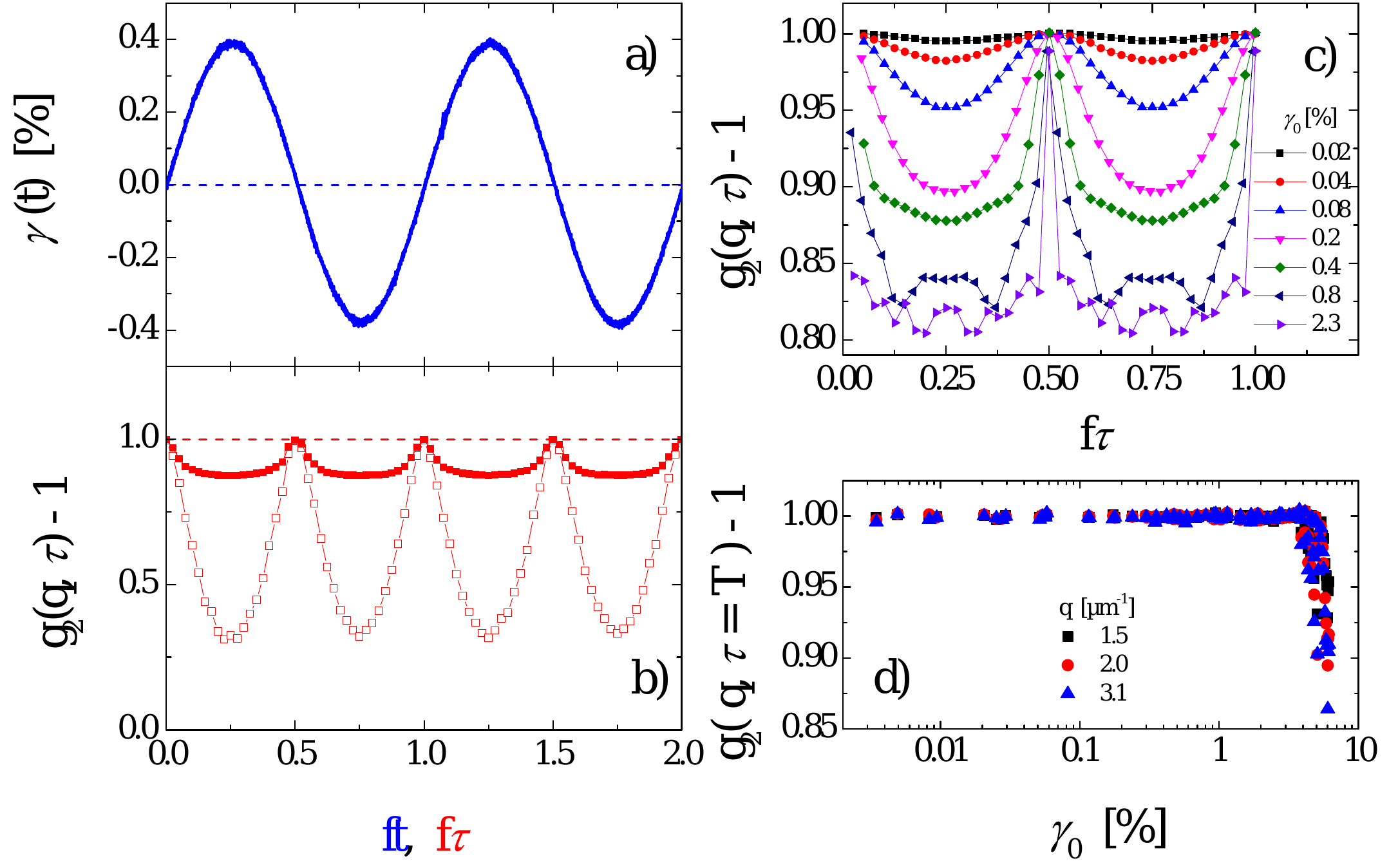}
  \caption{a) Oscillating strain under an oscillating stress of amplitude $\sigma_0/G_0=0.4\%$ as a function of adimensional time $ft$ ($f=0.025$~Hz is the frequency of the oscillation). b) Intensity correlation function measured for $q$ vectors oriented in the velocity direction (open symbols) and in the vorticity direction (filled symbols), respectively. Correlation echoes occur whenever the macroscopic deformation goes through 0. c) Correlation functions measured in the vorticity direction for various strain amplitudes $\gamma_0$, as shown by the label. In b) and c) the scattering vector is $q=3.1\um^{-1}$. d) Amplitude of the correlation echo as a function of the strain amplitude $\gamma_0$, for three representative scattering vectors. {\color {myc} Sample age is $t_w=1.5\times10^5$ s.}}
  \label{fig:revnonaff}
\end{figure}

	The reversibility of the imposed deformation is confirmed at a microscopic level by dynamic light scattering under oscillatory shear.
	By means of a two-time intensity correlation function (cf. Sec. \ref{sec:matmeth}), dynamic light scattering allows one to quantify the evolution of the sample microscopic configuration during deformation. Representative experimental results are shown in fig. \ref{fig:revnonaff}b, where the microscopic configuration of the sample at rest is compared with that of the sample under an oscillatory shear of amplitude $\gamma_0=0.4\%$ (fig. \ref{fig:revnonaff}a). Data for two wavevectors with equal magnitude $q=3.1\um^{-1}$ but different orientations, either parallel ($\vec{q}_\parallel$) or perpendicular ($\vec{q}_\perp$) to the shear direction, are compared. Both show a sudden drop as soon as the sample is deformed. This implies that the particles' relative position changes as a consequence of shear.
	
%------------------------------------------------
%------------------------------------------------
	
	We emphasize the fact that a significant decorrelation is also observed along $\vec{q}_\perp$, indicating that the sample deformation is not purely affine. Nonaffine deformations can be interpreted in two different ways. They could be due to plastic, irreversible rearrangements \cite{falk_dynamics_1998} or they may originate from spatial fluctuations of the sample elastic modulus \cite{basu_nonaffine_2011, didonna_nonaffine_2005, leonforte_inhomogeneous_2006}. Remarkably, we observe that for both components $\vec{q}_\parallel$ and $\vec{q}_\perp$ the correlation function increases again as the macroscopic deformation is reverted, and reaches values close to $1$ when $\gamma(t)=0$, e.g. for a time delay $\tau=1/f$, $f$ being the frequency of the oscillation: this behavior is referred to as \textit{correlation echo} and has been reported for a variety of soft systems under an oscillatory deformation of modest amplitude \cite{hebraud_yielding_1997, petekidis_rearrangements_2002, laurati_plastic_2014, rogers_echoes_2014}.
	The value of the correlation echo is very close to one, indicating that as the macroscopic deformation is recovered, the microscopic configuration reverts to that of the sample at rest. This unambiguously demonstrates  that no plastic events occur during the oscillatory shear in the linear regime (cf. fig. \ref{fig:revnonaff}c). In this regime, therefore, the non-affinities are due to spatial inhomogeneities of the sample elastic modulus, as reported for other network-forming systems \cite{basu_nonaffine_2011}. {\color {myc} Interestingly we note that non-affine reversible rearrangements have also been observed in the simulations of a gel in the linear regime and interpreted as a consequence of the extended floppy modes of the gel structure \cite{colombo_microscopic_2013, colombo_stress_2014}}. The scenario depicted in the linear regime changes as soon as $\gamma_0$ is increased beyond the linear regime: as shown in fig. \ref{fig:revnonaff}d, a sharp drop in the first correlation echo indicates the onset of irreversible rearrangements taking place in the nonlinear regime.
 This non-linear, irreversible regime is beyond the scope of this work. We simply mention that we have uncovered intriguing dynamical precursors of failure in the non-linear regime of creep tests \cite{aime_microscopic_2018-1}.

\section{Discussion and conclusion}
\label{sec:discussion}
	
	We have investigated in detail the linear viscoelasticity of a fractal colloidal gel. Thanks to a combination of several rheological tests, we have demonstrated that the linear viscoelasticity can be very well described by a Fractional Maxwell model. This model depends on three parameters, an elastic modulus, a characteristic time and an exponent, $\alpha$, {\color {myc} which remain yet to be related to some underlying physical processes. We note that a fractional Kevin-Voigt (FKV) model cannot account for our experimental data, since it predicts an increase of $G"$ with the frequency which is at odds with our measurements. Equivalently, one would not get an instantaneous elastic strain jump in creep with any Kelvin-Voigt-based model. \cite{bouzid_computing_2018}}
	
As mentioned in the introduction, an expression linking fractal dimension and power law rheology can be borrowed from the work of Muthukumar \cite{muthukumar_screening_1989} on critical gels. In the limit of screened hydrodynamic and excluded volume interactions, $\alpha$ is related to the fractal dimension $d_f$ through $\alpha=[3(5-2d_f)]/[2(5-d_f)]$. Using this expression, we find that the slight decrease of the exponent $\alpha$ of the FM model with sample age, from $0.45$ to $0.35$ (Fig.~\ref{fig:fmm_fitparams}) would correspond to a fractal dimension slightly increasing from $d_f=2.05$ to $2.15$. Within the experimental error, this prediction is in excellent agreement with the value of $d_f$ extracted from our scattering data (Fig.~\ref{fig:Iofq}). Although we {\color {myc} could not}  experimentally check the evolution of the fractal dimension with sample age as it requires a neutron beam, we note that a slow increase of $d_f$ with sample age has previously been measured in more dilute colloidal gels \cite{cipelletti_universal_2000}. The quantitative agreement with Muthukumar's prediction, further strengthens the connection between the fractal structure and the power law rheology, extending it beyond the framework of critical gels discussed in Ref.~\cite{muthukumar_screening_1989}. This is actually intriguing, since the original derivation \cite{muthukumar_screening_1989} was based on the assumption that the structure was scale-free, whereas in our case the fractal cluster size $\xi$ clearly emerges from Fig.~\ref{fig:Iofq} as a characteristic length scale of the system. {\color {myc} We note in addition that powerlaw rheology with a similar exponent $0.5$ could also be observed in numerical works for gels for which the exponent could not be straightforwardly related to a fractal structure. \cite{zia_micro-mechanical_2014, landrum_delayed_2016, bouzid_computing_2018}  Moreover possible differences between the static microstructure and the structure of the stress-bearing backbone might be an issue.}
	
	The presence of a cluster network of well defined connectivity challenges the validity of the FM model at the smallest frequencies. The power law distribution of relaxation times assumed by the model implies for instance that the creep deformation reaches arbitrarily large values at times long enough, whereas one should expect a saturation to a terminal plateau deformation at long times somewhere beyond $\tau_{FM}$. Unfortunately, such timescales are experimentally inaccessible, since the characteristic time grows linearly with sample age: as a consequence, any experiment probing a timescale beyond $\tau_{FM}$ would automatically be affected by aging. For this reason, the only way to test the consistency of FM at very small frequencies would be to repeat this analysis on different samples, for example changing the volume fraction. Based on a simple analysis with the time-cure superposition for polymer gels {\color {myc} (see e.g.~\cite{adolf_time-cure_1990}}) and discussed in the introduction,
one might expect that, with increasing volume fraction, the sample would be increasingly far away from the critical gel point: the range of length scales displaying self-similarity would be thus reduced and consequently one might expect an eventual terminal plateau to be shifted at higher frequencies.
	
	On the other hand, we argue that the impossibility of addressing the putative loss peak around $\tau_{FM}$ in any clean experiment questions its physical relevance. The usual interpretation of a fractional rheological model is based on the existence of a power law distribution of relaxation times, of which $\tau_{FM}$ should represent some average value. However, one could argue about what physical interpretation should be attributed to relaxation times beyond the characteristic time for physical aging.
	Indeed, physical aging represents a true challenge for the fractional calculus-based framework presented in this paper, since it has to be introduced `by hand' in the model in order to account for the experimental data. Therefore, one may wonder whether a more natural framework could exist that accounts for both the power law rheology and its time dependence.
	One potential candidate is Soft Glassy Rheology (SGR) \cite{sollich_rheology_1997}, which considers that the sample mechanics is controlled by disorder, metastability and local structural rearrangements.
	The advantage of SGR is that aging emerges as a natural consequence of these features, and SGR predicts a virtual structural relaxation time linearly increasing with sample age \cite{fielding_aging_2000}, compatible with our findings. Within this framework, our gel would be represented by an effective `noise temperature' $x=1-\alpha\sim 0.6$, well below the glass transition (which in SGR occurs at $x=1$). The frequency spectrum of the sample is then predicted to be characterized by a nearly constant elastic modulus $G'$ and a loss modulus $G''\sim \omega^{x-1}$ \cite{sollich_rheological_1998}, which is in good agreement with the data in the experimentally accessible frequency range (Fig.~\ref{fig:dfs_wide}).
	However, one feature that appears to be in stark contrast with the picture offered by SGR is the complete reversibility of the deformation, as demonstrated by the creep recovery experiments (see e.g. Fig.~\ref{fig:creep_and_recovery}) and by the dynamic light scattering coupled to shear (Fig.~\ref{fig:revnonaff}). By contrast, in SGR creep is never completely recoverable, and the non-recoverable amount depends on the creep duration.
Moreover, SGR predicts a logarithmic creep in the glassy regime ($x<1$) \cite{fielding_aging_2000}, which is clearly incompatible with our experimental data (Fig.~\ref{fig:aging_dfs}).
	
	On the one hand, these considerations explicitly demonstrate that a thorough comparison between different rheological experiments allows one to discriminate between models that would be indistinguishable only looking at one kind of test, say oscillatory rheology. On the other hand, they raise the question of whether some concepts of SGR could be borrowed to provide a more natural framework to introduce aging in fractional models.
	This possibility opens intriguing perspectives for future theoretical works, aiming at a deeper understanding of the microscopic origin of power law rheology.

{\color {myc} In addition, we have shown that non-affine but reversible rearrangements occur in the linear regime, thus hinting at a possible connection between reversible non-affinities and power law rheology. Relating those non affinities to some physical parameters as done in \cite{basu_nonaffine_2011} is delicate and would certainly deserve deeper investigation, by performing for instance spatially resolved measurements.}

\begin{acknowledgments}

This work was supported by the EU (Marie Sklodowska-Curie ITN Supolen, Grant No. 607937) and ANR (Grant No. ANR-14-CE32-0005-01).
We are grateful to LLB for beam time and to Dafne Musino and Justine Pincemaille for the neutron scattering measurements. We also thank Thibaut Divoux for fruitful discussions.

\end{acknowledgments}

\end{document}